\documentclass[12pt,preprint]{aastex}
\usepackage{amsmath}
\usepackage{lscape}
\newcommand{\degree}{\ensuremath{^\circ}\/}

\shorttitle{
Polar Field Reversal Observations with {\it Hinode} 
}
\shortauthors{Shiota, et al.}
\begin{document}

\title{Polar Field Reversal Observations with {\it Hinode}}

\author{D. Shiota\altaffilmark{1}, 
S. Tsuneta\altaffilmark{2}, M. Shimojo\altaffilmark{2}, N. Sako\altaffilmark{3}, 
D. Orozco Su{\'a}rez\altaffilmark{2, 4}, R. Ishikawa\altaffilmark{2}
}
\altaffiltext{1}{Advanced Science Institute, 
RIKEN (Institute of Physics and Chemical Research), 
Wako, Saitama 351-0198 Japan;
shiota@riken.jp
}
\altaffiltext{2}{National Astronomical Observatory of Japan(NAOJ), Mitaka, Tokyo 181-8588  Japan} 
\altaffiltext{3}{The Graduate University for Advanced Studies, Mitaka, Tokyo 181-8588, Japan}
\altaffiltext{4}{Instituto de Astrofisica de Canarias, 38205, La Laguna, Tenerife, Spain}

\begin{abstract}
We have been monitoring yearly variation in the Sun's polar magnetic fields with the Solar Optical Telescope aboard {\it Hinode} to record their evolution and expected reversal near the solar maximum. All magnetic patches in the magnetic flux maps are automatically identified  to obtain the number density and magnetic flux density as a function of th total magnetic flux per patch. The detected magnetic flux per patch ranges over four orders of magnitude ($10^{15}$ -- $10^{20}$ Mx). The higher end of the magnetic flux in the polar regions is about one order of magnitude larger than that of the quiet Sun, and nearly that of pores. Almost all large patches ($ \geq 10^{18}$ Mx) have the same polarity, while smaller patches have a fair balance of both polarities. The polarity of the polar region as a whole is consequently determined only by the large magnetic concentrations. A clear decrease in the net flux of the polar region is detected in the slow rising phase of the current solar cycle. The decrease is more rapid in the north polar region than in the south. The decrease in the net flux is caused by a decrease in the number and size of the large flux concentrations as well as the appearance of patches with opposite polarity at lower latitudes. In contrast, we do not see temporal change in the magnetic flux associated with the smaller patches ($ < 10^{18}$ Mx) and that of the horizontal magnetic fields during the years 2008--2012. 
\end{abstract}

\section{INTRODUCTION}
Observations of the magnetic field in the Sun's polar region are of crucial importance to understand the long-term variation of solar magnetism \citep{2011IAUS..273...28C}. Each polar region was believed to be covered by extended weak unipolar magnetic fields sometimes referred to as ``polar caps,'' for several years around the solar minimum. The pioneering observations of polar faculae \citep{2004A&A...425..321O, 2007A&A...474..251B} provided new and important information on the magnetism of the polar regions. However, it is the global magnetic landscape of the polar region obtained with  {\it Hinode} that really changed our paradigm for the polar magnetic fields as described below \citep{2008ApJ...688.1374T}.  The magnetic polarity of the polar regions is known to be reversed at around the solar maximum \citep{1959ApJ...130..364B}. Since the start of the new solar cycle in December 2008 \citep{sidc},  solar activity has been rising, and the polarity reversal is expected to occur by the middle of 2013.

Current solar dynamo models provide us with the following scenario for  polar field reversal (\citealt{1991ApJ...383..431W}; \citealt{1995A&A...303L..29C}; \citealt{2004ApJ...601.1136D}): Magnetic fields with minor polarity (polarity opposite to the dominant polarity of the polar region) are transported from active region remnants at mid-latitudes by meridional flow and turbulent diffusion \citep{2010Sci...327.1350H}. The local magnetic flux in the polar region is apparently canceled by the incoming opposite magnetic flux somewhere near the polar cap boundary. As seen in the measurements of the line-of-sight magnetic component (\citealt{1988assu.book.....Z, 1998SoPh..177..375F, 2000SoPh..191....1S, 2004A&A...428L...5B}), the polar cap boundary retracts poleward. Thus as increasingly more magnetic flux arrives from lower latitudes, the polar regions will again be dominated by unipolar flux, but this time with opposite polarity. 

The actual process of reversal is, however, still poorly understood, partly because of the apparent difficulties in conducting observations of the polar region from the ground, and specifically the foreshortening and a strong intensity gradient combined with variable  seeing \citep{1996SoPh..163..223L, 2004A&A...425..321O}. The high-resolution and high-sensitivity spectropolarimetric observations with {\it Hinode} almost completely alleviated the problems associated with polar field measurements, and provide us with a clear picture of the magnetic landscape of the polar region for the first time \citep{2008ApJ...688.1374T}.  

The spectropolarimeter  (SP;  \citealt{2001ASPC..236...33L}) of the Solar Optical Telescope (SOT \citealt{2008SoPh..249..167T, 2008SoPh..249..197S, 2008SoPh..249..221S, 2008SoPh..249..233I}) aboard the {\it Hinode} satellite \citep{2007SoPh..243....3K} allows us to perform diffraction-limited, high-polarization-sensitivity observations, which reveal the fine structure of photospheric vector magnetic fields. Indeed, the first  {\it Hinode} SOT observations of the polar areas revealed the existence of many patchy magnetic concentrations with intrinsic field strengths of over 1 kG distributed across the entire polar region \citep{2008ApJ...688.1374T}. These flux concentrations have a form of unipolar strong (kilogauss order) fields. \cite{2010ApJ...719..131I} presented polar maps of the vertical and horizontal magnetic fields.

The existence of ubiquitous horizontal (highly inclined) magnetic fields in the quiet Sun was reported by \cite{2007ApJ...670L..61O} and \cite{2008ApJ...672.1237L}. These horizontal magnetic fields are smaller than granules, and are transient with lifetimes ranging from 1 to 10 min. Furthermore, the intrinsic magnetic fields are essentially weaker than the equipartition fields associated with granular convection \citep{2009A&A...495..607I}. There is no difference in the histograms of the intrinsic magnetic field strengths between the quiet Sun and the plage regions \citep{2009A&A...495..607I} and between the north polar region and the quiet Sun \citep{2010ApJ...719..131I}. These observations suggest that the transient horizontal magnetic fields result from the dynamo amplification of seed magnetic fields with the surface turbulent convection. 

It is important to examine how the horizontal and vertical magnetic fields permeating the polar regions change with and during the polarity reversal process. High-resolution, high-polarimetric-sensitivity vector magnetic observations with {\it Hinode} are indeed ideal for this purpose. We have been regularly visiting the north and south polar regions during the {\it Hinode} Observation Program (HOP) No. 2, 81, and 206. This paper presents the changing magnetic landscape of the polar regions with the advent of a new solar cycle.

\section{OBSERVATIONS AND DATA ANALYSIS} 

The rotation axis of the Sun is tilted by 7\degree\ with respect to the ecliptic plane. Thus, by choosing the proper timing, we are able to see the deep polar region as well as the pole itself. We specifically choose the September time frame for the north pole observations, and the March time frame for the south pole observations. The actual timing of the observations is usually affected by other observing programs. This fluctuation in observing timing in turn slightly affects the observed latitudinal range. We have examined several datasets of the polar regions collected from 2007 to 2012 with SP to identify those that are appropriate for comparison. Detailed information of the {\it Hinode} data analyzed in the present analysis is tabulated in Table 1. The dataset consists of snapshot images of the polar regions, each acquired through  several hours of slit scan observations. 

The {\it Hinode}/SP recorded the Stokes profiles of the \ion{Fe}{1} 630.15 and 630.25~nm spectral lines with a spatial resolution of about 0\farcs32 (0\farcs16 pixel size). The exposure time was 12.8~s for each slit position (except for the south pole dataset taken in 2007, where the exposure time was 4.8~s). We inferred the magnetic field vector and its fill fraction in each of the pixels by applying a least-square fitting to the observed Stokes profiles. We used the Milne-Eddington Inversion of Polarized Spectra (MILOS) code \citep{2007A&A...462.1137O}. The 10 free parameters are the three components describing the vector magnetic field (strength $B$, inclination angle, and azimuth angle), the line-of-sight velocity, two parameters describing the source function, the ratio of line to continuum absorption coefficients, the Doppler width, the damping parameter, and the stray-light factor $\alpha$. 

To obtain reliable information about the magnetic field vector, we excluded the pixels whose circular or linear polarization signal amplitudes are less than five times the averaged noise level $\sigma$ for each dataset. The noise level  $\sigma$ was determined in the continuum wavelength range of the Stokes V profiles, and was estimated to be  $\sigma/I_c \sim 0.001$ for the north polar region, where $I_c$ is the continuum intensity.

Once the data were inverted with the MILOS code, we resolved the 180\degree\ ambiguity for the transverse magnetic fields through the approach described in \cite{2010ApJ...719..131I}. First, two solutions of the vector magnetic field (due to the 180\degree\ ambiguity) for each pixel are transformed into those with respect to the local reference frame, and the local zenith angles of the two solutions are calculated. A key feature of the analysis is the validity of the assumption \citep{2007ApJ...670L..61O} that the magnetic field vector is either vertical or horizontal to the local surface (or undetermined) to resolve the 180\degree\ ambiguity. The zenith angle is defined to be from 0\degree\ to 40\degree\ and from 140\degree\ to 180\degree\ for the vertical magnetic field. The zenith angle of the horizontal field is defined to be between 70\degree\ and 110\degree\, following \cite{2009A&A...495..607I}. According to this definition, every pixel has a magnetic field vector classified as either vertical or horizontal (or undetermined).

Next, each magnetic field vector map is transformed into polar maps; that is, we display the data as it would be seen from just above each pole (see Figure 1). To finish, the intrinsic magnetic field ($B$)
is converted into the magnetic flux, i.e., the intrinsic magnetic field is multiplied by the magnetic filling factor ($f=1-\alpha$), by the cosine of the zenith angle, and by the pixel area, in the case of vertical fields. In the case of horizontal field data, they are multiplied by the sine of the zenith angle and by the square root of the pixel area times a geometrical depth of 190~km (see Section 6.3 in \citealt{2010ApJ...713.1310I}).

With this procedure, we obtained maps of the magnetic flux for the magnetic field vectors classified as either vertical or horizontal.  Figure 1 shows polar views of the magnetic flux of the vertical magnetic vectors in the north polar region on 2008 September 20 and 2011 October 9. We now know that strong intrinsic magnetic fields  are concentrated in the large patchy areas seen in the maps \citep{2008ApJ...688.1374T}. In order to obtain information on the magnetic structures, we clustered neighboring pixels with the same polarity into a single group, and identified and treated each group as a single flux concentration. The circles in Figure 1 show the location of such flux concentrations whose total magnetic flux is larger than 10$^{18}$Mx. Here the total magnetic flux was calculated by summing the flux of all pixels within each circle.

This clustering procedure was applied to the magnetic fields classified as vertical and horizontal and located at latitudes higher than 70\degree. We calculated the number of magnetic patches as a function of the total magnetic flux. This is referred to as the number density histogram in this paper.  We also calculated the average magnetic flux density provided by all the magnetic patches within a given magnetic flux range (flux density histogram). The average flux density is defined to be the total flux  (contributed by the magnetic concentrations within a given range of magnetic flux) divided by the SOT observing area  (70\degree\ - 90\degree\ latitudes). (We set the boundary of 75\degree\ latitude for the data taken in 2007 in order to directly compare the histograms with that of the quiet Sun observations in Figure 2.)

We can infer how much of the average flux density of the entire polar region is contributed by the magnetic patches within a specific range of magnetic flux. For instance, Figure 2(d) indicates that magnetic patches with fluxes from ${3.5 \times 10^{18}}$ Mx to ${10^{19}}$ Mx contribute about 0.26 Mx cm$^{-2}$ to the total average flux density (1.24 Mx cm$^{-2}$) of the north polar region. 

We point out that the clustering approach to treat each flux concentration as a single entity may better elucidate their contribution to the global magnetic field in the polar regions: simple pixel-based histograms \citep{2008ApJ...688.1374T, 2010ApJ...719..131I} do not allow us to obtain information on the relative contribution of magnetic patches of various sizes to the global polar magnetic fields and their respective time evolution. 

\section{PROPERTIES OF POLAR MAGNETIC FIELDS}

\subsection{Properties of Vertical Magnetic Fields in Polar Regions}
In this section, we compare the histograms of flux concentrations in the polar regions and the quiet Sun at the East limb. The quiet Sun data is included in the analysis for reference. The observations of the quiet Sun are described in detail in \cite{2010ApJ...719..131I}. We confirmed that there is no coronal activity (enhanced brightness) in the region on the basis of the data from the X-ray telescope aboard {\it Hinode}.  

The top panels in Figure 2 show the number density of the flux concentrations as a function of their total magnetic flux for vertical magnetic fields. The detected magnetic flux per patch ranges over four orders of magnitude from 10$^{15}$ Mx to 10$^{20}$ Mx for polar regions, while the maximum magnetic flux is approximately 10$^{19}$ Mx for the case of the quiet Sun. The larger end of the magnetic flux in the polar regions is comparable to that of pores ($2 \times 10^{19} \sim 1 \times 10^{20}$ Mx) \citep{1998ApJ...507..454L, 2012ApJ...747L..18S}. 

The number density decreases with increasing clustering flux for more than four orders of magnitude. The three regions (polar regions and the quiet Sun) have comparable number densities for both polarities in the range from $10^{15}$~Mx to $10^{17}$~Mx per concentration, while there is a significant imbalance in the number density for flux concentrations above $10^{17}$~Mx in the polar regions. The imbalance becomes more prominent for larger magnetic flux (per concentration).
 
The bottom panels in Figure 2 show histograms of magnetic concentrations in terms of the average flux density. We clearly see the same property: one polarity dominates toward larger clustering flux. In particular, the flux concentrations with total flux $ \geq 10^{18}$~Mx primarily determine the polarity and the total magnetic flux of the polar regions: negative for the north pole and positive for the south pole. The weaker patches with both polarities tend to cancel out. However, the positive and negative patches in the quiet region are remarkably balanced across the entire flux range.

\subsection{Long-Term Variation of Polar Magnetic Fields}
Figure 3 compares the histograms during the solar minimum, i.e., 2008 and subsequent years of rising activity until 2012. All the data shown in Figure 3 are taken with the same exposure time, allowing direct comparison. The signed average flux density decreases with time. The signed magnetic flux density of the north polar region decreases much faster than that of the south polar region. Note that the magnetic flux contribution of the large flux range ($ \geq 10^{18}$~Mx) clearly decreases, while that of the small flux range (between $10^{15}$~Mx and $10^{17}$~Mx) essentially stays the same. As described in Appendix A, we detected a slight degradation in the instrument throughput (10\% over three years). This should not, however, affect these observations due to our stringent $5\sigma$ threshold for pixel selection (see Appendix A for details).

As the number of large concentrations of the dominant polarity decreases, their areas decrease as well. This tendency is clearly seen in Figure 1: the number of large concentrations in the map for 2011 October 9 is smaller than that for 2008 September 20.
 
We also calculated the histograms of horizontal flux concentrations using the same procedure. The average flux density for three years is shown in Figure 4. The shape of the histograms is quite different from that of the vertical magnetic fields. They have a peak at around $10^{18}$ Mx. In contrast to the significant variation for the vertical magnetic fields, the variation is relatively small ($\pm 8 $\%) in this case. The shapes of the histograms appear to be maintained during the three-year period. 

To better illustrate the temporal variation seen in Figures 3 and 4, we examine the total flux density in Figure 5. Figures 5(a) and 5(b) show the total flux density of the vertical magnetic fields contributed by the large ($\geq 10^{18}$~Mx) magnetic concentrations, while Figures 5(c) and 5(d) show the total flux density contributed by the small concentrations ($<10^{18}$~Mx). The magnetic flux provided by the large flux concentrations rapidly decreases in the north polar region (Figure 5(a)), while the decrease is much less significant in the south polar region (Figure 5(b)). In contrast, the contribution to the total flux density from the patches with smaller magnetic flux remains comparable and stable in the polar regions (Figures 5(c) and 5(d)). Finally, we note that although the total flux density of the horizontal magnetic fields somewhat fluctuates in both regions, this fluctuation does not seem to have a correlation with the vertical magnetic flux (Figure 6).
 
The boundary between the large and small flux concentrations is located at around  $10^{18}$~Mx as mentioned above. The critical flux agrees with that separating the quiet Sun network from the internetwork fields \citep{1995SoPh..160..277W}. The mixed polarity nature of the weak vertical magnetic fields is consistent with the properties of the quiet Sun internetwork fields, while the large flux concentrations of the vertical magnetic fields do not have an apparent counterpart in the quiet Sun.

We notice a short-term variation in data collected several days apart (Table 1) in Figure 5. The difference between the pair of nearby measurements appears to be larger for the large magnetic concentrations of the vertical magnetic fields and the horizontal magnetic fields, while that of the small magnetic concentrations appears to be small. The large magnetic concentrations have a lifetime of about 10 h, and only several tens of such patches (Figure 1) essentially determine the global polar fields. This situation may introduce statistical fluctuation as far as the large magnetic concentrations are concerned, or these fluctuations may be of an intrinsic nature, though we do not rule out the effect of photon statistics.

\section{SUMMARY AND DISCUSSION}

\subsection{Two Magnetic Components in Polar Regions}
We have presented histograms of the magnetic flux per patch in both polar regions and in the quiet Sun. For the vertical magnetic fields, significant imbalance between the positive and negative polarities was observed mainly at large flux concentrations ($ \geq10^{18}$~Mx) in both polar regions, whereas small flux concentrations have mixed, well-balanced polarities. The quiet Sun does not show polarity imbalance at any flux scale. 

\cite{2011ApJ...737...52L} reported that positive and negative polarities of the internetwork magnetic fields are well balanced, while the network magnetic fields have a considerable polarity imbalance. In our observations of the quiet Sun (Figures 2s (c) and 2 (f)), both the network and internetwork fields are included in the histograms. The lower end of the magnetic flux distribution would correspond to the internetwork fields, and both results are consistent. The higher end of the magnetic flux distribution would correspond to the network fields, and our result is different from that of \cite{2011ApJ...737...52L}. Depending on the location and the timing of the observations, the degree of flux imbalance may differ owing to differences in the environment, such as the presence of nearby active region remnants.

It is remarkable that polar regions are able to create magnetic concentrations with magnetic flux as high as that of pores. The large magnetic concentrations are seen as bright polar faculae \citep{2008ApJ...688.1374T} in white light, while pores with the same magnetic flux are seen as dark features at the disk center. The significant difference in the appearance of the two flux tubes may be related to the viewing angle as well as the formation and structure of polar magnetic concentrations and pores. 

The high end of the magnetic flux is about one order of magnitude higher than that of the quiet Sun (Figure 2). There may be a higher chance in the quiet Sun for the positive and negative patches to collide, reconnect, and lose magnetic energy or submerge as a result. As such, the environment in the quiet Sun may not allow the elemental magnetic concentrations to grow. In contrast, polar regions have a dominant polarity as a whole, and positive and negative magnetic fluxes are not balanced. This allows magnetic concentrations to grow in the favorable environment of like polarity \citep{2010ApJ...719..131I}. 

We also showed, for the first time, the yearly variation in the polar magnetic fields. The net average flux density in the polar regions decreases with time. The decrease in the north polar region is faster than that in the south polar region. The decrease is mainly caused by the reduction in the number and size of the large concentrations of the dominant polarity. Note that the polarity reversal is not yet complete even at the north pole. However, if the decreasing rate of the average signed flux density in the north polar region ($ \sim0.28 $ Mx cm$^{-2}$ yr$^{-1}$) continues, the reversal is expected to complete by the middle of 2012.

The average magnetic flux density of the small vertical flux concentrations ($ 10^{15} \sim 10^{17}$~Mx) in polar regions has been stable in the present rising phase of solar activity. The average magnetic flux density of the horizontal magnetic fields has been stable as well.

The above results clearly indicate that the magnetic concentrations of the vertical magnetic fields in the polar regions consist of two different components. One corresponds to the large flux concentrations (large in terms of total magnetic flux per patch) with a dominant polarity. The total flux density contributed by this component varies with the solar cycle. The other is composed of the small flux concentrations of both polarities (small in terms of total unsigned magnetic flux per patch). The total flux density of this component does not vary with the solar cycle. 

\subsection{Horizontal Magnetic Fields and Origin of Small Magnetic Concentrations}
As discussed in Section 1, there is strong evidence from {\it Hinode} suggesting that the horizontal magnetic fields are generated by a local dynamo process. This is further strengthened by the present observations that the average flux densities of the horizontal magnetic fields in both polar regions are almost the same, and that they did not change with time from 2008 to 2012, when the average flux density of the vertical magnetic fields in the north polar region changed considerably. 

There is a growing body of evidence indicating that some fraction of the transient horizontal magnetic field evolves into bipolar vertical fields in the internetwork region. \cite{2009ApJ...700.1391M} reported that 23\% of the small-scale horizontal magnetic fields that emerge in the photosphere reach the chromosphere, leaving behind the bipolar footpoints. \cite{2010ApJ...713.1310I} clearly showed using Stokes Inversion based on Response function (SIR) analysis that the horizontal magnetic fields rise up through the photosphere reaching the layers above the line-forming region of \ion{Fe}{1} 630.2 nm lines, and the horizontal field eventually disappears, leaving the bipolar vertical magnetic fields. Furthermore, \cite{2011ApJ...735...74I} discovered a clear positional association between the vertical and horizontal magnetic fields in the internetwork region. All of the horizontal magnetic patches are associated with the vertical magnetic patches. They conjecture that internetwork magnetic fields are formed by the emergence of small-scale flux tubes with bipolar footpoints, and the vertical magnetic fields of the footpoints are intensified to kilogauss fields via convective collapse. 

On the basis of this conjecture, we herein compared the observed average flux density of the small concentrations with the estimated values from these prior {\it Hinode} observations of the horizontal magnetic fields. The vertical magnetic flux supplied by the emergence of the horizontal magnetic fields was estimated to be $ 0.2$~Mx~cm$^{-2}$ with the following typical observed parameters for the horizontal fields: magnetic flux carried by individual transient horizontal magnetic fields is  $ \sim 10^{17}$~Mx, and the occurrence rate is  $\sim 10^{-20}$~cm$^{-2}$~s$^{-1}$ \citep{2009A&A...495..607I}. The efficiency of emergence to the chromosphere and possibly to the corona is 23~\% \citep{2009ApJ...700.1391M}. The lifetime of the vertical magnetic fields is 750~s \citep{2009ApJ...700.1391M}. The estimated vertical magnetic flux is comparable with the observed magnetic flux for the small flux concentrations (approximately $0.3$~Mx~cm$^{-2}$  from Figure 5(c) and 5(d)). This may support the scenario that the vertical magnetic fields with mixed polarities are locally generated in a form of the horizontal magnetic fields, as suggested by \cite{2011ApJ...735...74I}. 
 
\subsection{Origin of Large Flux Concentrations}
Our observations reveal that two different magnetic structures coexist in the polar regions. One clearly represents the single-polarity, large flux concentrations  that vary with the solar cycle, and the other corresponds to small flux concentrations whose flux seems to be maintained during the whole solar cycle by a local process. Polarity reversal would be due to the variation in the distribution of the large flux concentrations. 

The generation and maintenance of such large flux concentrations remains an open question. One crucial observation \citep{2008ApJ...688.1374T} is that the vertical kilogauss patches have relatively short lifetimes of 5 to 15 h: such large patches appear alone or are formed as a result of coalescence of multiple smaller patches. The patches thus created split into multiple smaller patches and/or disappear alone, while maintaining unipolarity. This unipolar appearance and disappearance suggest that the large patches are formed from and disintegrated into patches with magnetic flux below the detection limit of the instrument. The component seen in the $10^{15} \sim 10^{16}$~Mx~cm$^{-2}$ range may be the tip of the iceberg of these unseen fluxes, and the large concentrations may have formed from the inventory of small concentrations.

Figure 3 shows that the positive and negative fluxes are well balanced at the lower end of the magnetic flux. Imbalance, however, starts to appear and becomes large with increasing magnetic flux per patch. This, together with the above observations, indicates that the small component is not completely independent of the large component, and is somehow affected by the large component or vice versa. 

\subsection{Polar Field Reversal}
Figure 5(a) shows that the decreasing rate of the magnetic flux in the north polar region slowed down slightly or was at most stationary from 2008 to 2012. The reversal of the polar field is supposed to take place near the solar maximum, which is associated with the eruption of many active regions. However, this minimum between 2008 and 2010 is characterized by a long-lasting dormancy of sunspot emergence. Even in this situation without apparent transport of magnetic flux of opposite polarity, the magnetic flux in the north polar region continue to decrease, while that of the south polar region remains constant. This may not be consistent with the standard flux transport dynamo model, which requires active regions that provide magnetic flux to reverse the polar fields. 

In Figure 1(b), it can be seen that the magnetic flux of opposite polarity appears at the lower latitudes of 70\degree - 75\degree\ in a later year (2011). This suggests that the opposite polarity flux is transported to the polar region via meridional flow and/or a turbulent diffusion process as the solar activity increases. Observational clarification of the relationship among the large unipolar flux concentrations, the transported opposite magnetic flux from lower latitudes, and the small bipolar concentrations and the horizontal magnetic fields in polar regions would be of crucial importance to understand the full inventories of solar magnetism and the solar dynamo process. 

\acknowledgments
{\it Hinode} is a Japanese mission developed and launched by ISAS/JAXA, collaboration with domestic partner NAOJ and international partners NASA and STFC (UK). Scientific operation of the {\it Hinode} mission is conducted by the {\it Hinode} science team at ISAS/JAXA. This team mainly consists of scientists from institutes in the partner countries. Support for the post-launch operation is provided by JAXA and NAOJ, STFC, NASA, ESA, and NSC (Norway). D.S. is supported by the Special Postdoctoral Researchers Program at RIKEN.

\appendix
\section{LONG-TERM THROUGHPUT VARIATION OF SOT/SP}
It is important to check for possible variation in the instrument throughput for such a synoptic study. The throughput of the entire SOT/SP instrument depends on the throughput of the optical telescope assembly (OTA) multiplied by that of the spectropolarimeter (SP) of the focal plane package (FPP). Figure \ref{Icont} shows the long-term variation of the Stokes I continuum intensity from 2008 to 2012. The data set used in the figure is the same as that used in the analysis. The Stokes I continuum intensity is averaged over a specific region defined as follows: The size of the box is 32 arcsec in the $x$ direction (solar west) and 8 arcsec in the $y$ direction (solar north), and its center is located at $x$ = 0 and 72.8 arcsec inside the limb in the $y$ direction. Pixels with high polarization signal ($\geq 5\sigma$) are removed. Averaging is also done over the wavelength range 6300.892 -- 6300.999 \AA, which is outside the \ion{Fe}{1} absorption lines. 

Figure \ref{Icont} clearly shows that the throughput of SOT/SP gradually decreases by about 10\% over three years. The rate of decrease in throughput appears to have become smaller with time. As a result, the polarization sensitivity of the SOT/SP instrument is slightly compromised by about 5\% in 2012 with respect to that in 2008, simply because the photon noise $\sigma$ is proportional to the square root of the Stokes I intensity. 

The reason why we have a stringent criterion, viz., $5\sigma$ for valid pixels is to avoid contamination with false polarization signals that stem from the photon noise (and possible cosmic ray hits). The threshold is located above from level of the photon noise, and any small change in throughput does not affect the detection capability of the magnetic flux concentrations. In particular, the large flux concentrations emit intense polarization signals, and the small change in throughput does not affect their detectability.



\begin{figure}
\begin{center}
\includegraphics[height=15cm]{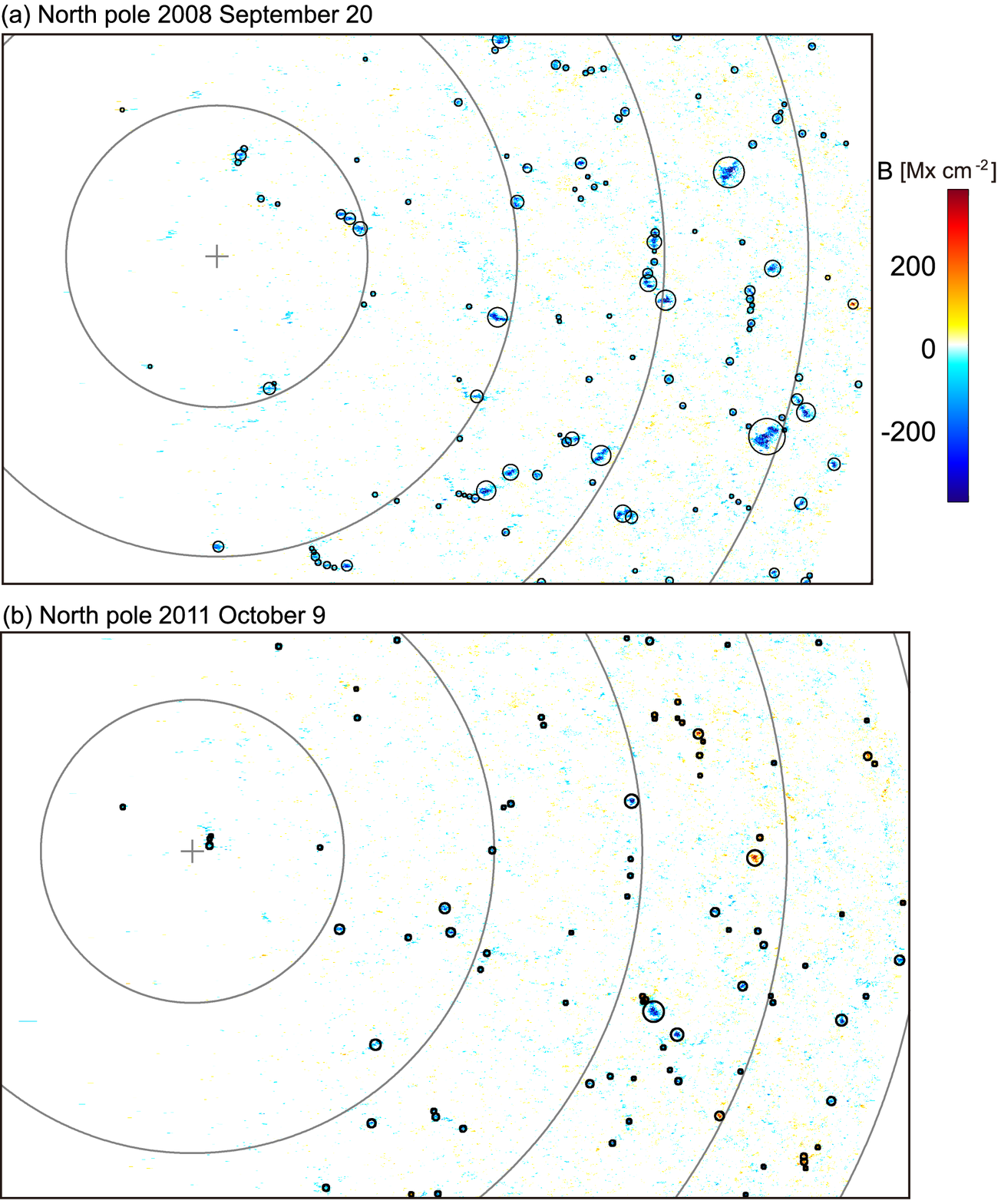}
\end{center}
\caption{
Maps of signed magnetic flux of the magnetic field vectors classified as vertical in the north polar region on 2008 September 20 (top) and on 2011 October 9 (bottom) (see \cite{2010ApJ...719..131I} for the methodology). The exposure times of the two images are the same (12.8~s). The crosses indicate the north pole. Bold lines show longitude arcs separated by 5\degree. Small circles indicate the automatically identified patches with total magnetic flux per patch above $10^{18}$ Mx. 
The circle radii are proportional to the squares root of the total magnetic flux of the patches.}
\label{map}
\end{figure}

\begin{figure}
\begin{center}
\includegraphics[width=15cm]{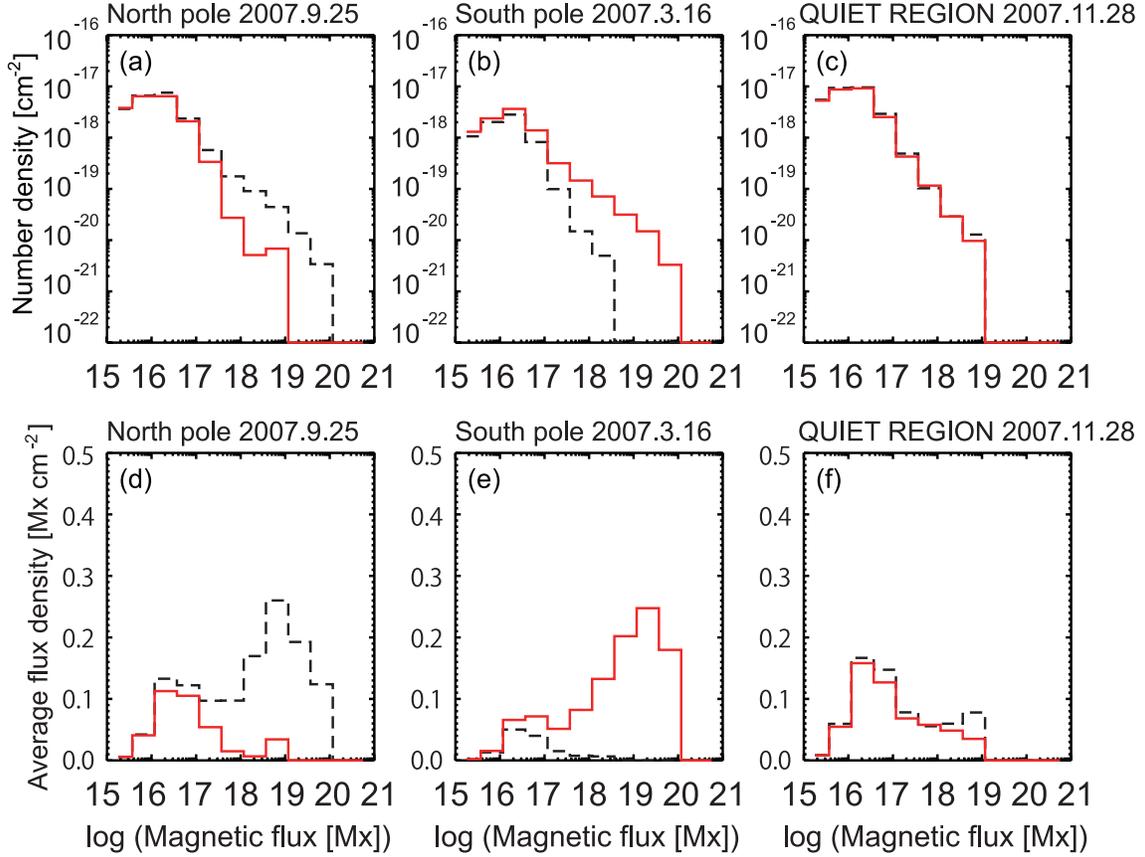}
\end{center}
\caption{
Histograms of magnetic flux per patch in terms of number density (top panels) and average flux density (bottom panels) for north pole (left), south pole (center), and the quiet Sun region at the East limb (right). The magnetic flux concentrations here refer to the concentrations of the vertical magnetic vectors. The average flux density is defined as the total flux  (contributed by the magnetic concentrations in each bin of the horizontal axis) divided by the SOT observation area.  Black dashed  and red bold lines represent the negative and positive concentrations, respectively. The data for the north and the south polar regions and the quiet region were obtained during 00:10 -- 07:26 UT on 2007 September 25 (panels a and d), 12:02 -- 14:56 UT on 2007 March 16 (panels b and e), and 18:36 -- 20:52 UT on 2007 November 28 (panels c and f), respectively. The exposure times for the north polar region and the quiet Sun are the same (12.8~s).  The exposure time for the south polar region is shorter (4.8~s); thus, that the south pole histogram cannot be directly compared with those for the north pole and the quiet Sun.
}
\label{his}
\end{figure}

\begin{figure}
\begin{center}
\includegraphics[width=15cm]{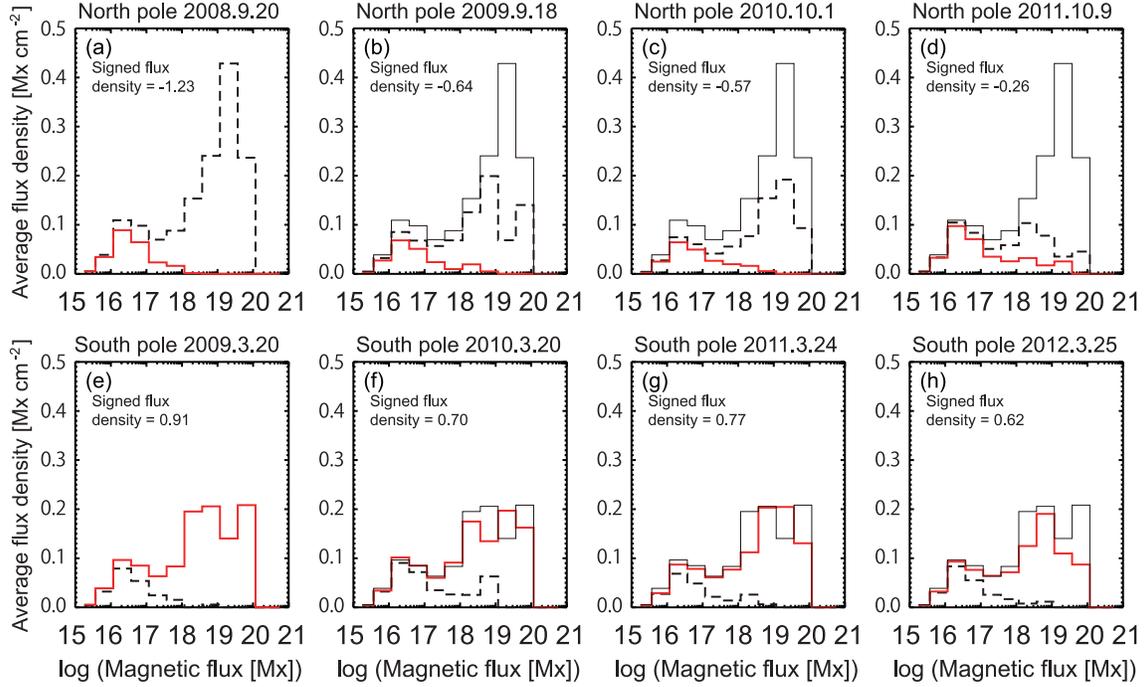}
\end{center}
\caption{
Histograms of magnetic flux density per concentration in the north and south polar regions from 2008 to 2012. The magnetic flux concentrations here refer to the concentrations of the vertical magnetic vectors. The average flux density is defined as the total flux  (contributed by the magnetic concentrations in each bin of the horizontal axis) divided by the SOT observation area. Total (signed) average flux density is tabulated in each panel.   Bold red and dashed black lines represent the negative and positive vertical magnetic vectors, respectively. North pole data were obtained on
(a) 2008 September 20,
(b) 2009 September 18,
(c) 2010 October 1,
and (d) 2011 October 9 (N1, N4, N5, and N8 in Table 1, respectively).
South pole data were obtained on
(e) 2009 March 20,
(f) 2010 March 20,
(g) 2011 March 24,
and (h) 2012 March 25 (S2, S3, S6, and S7 in Table 1, respectively). 
The exposure times for these data are the same (12.8 s).  The black thin lines in panels (b), (c), and (d) are the same as the negative polarity histogram (dashed black line) in panel (a), and those in panels (f), (g), and (h) are the same as the positive polarity histogram (bold red line) in panel (e) for comparison. 
}
\label{yv}
\end{figure}

\begin{figure}
\begin{center}
\includegraphics[width=15cm]{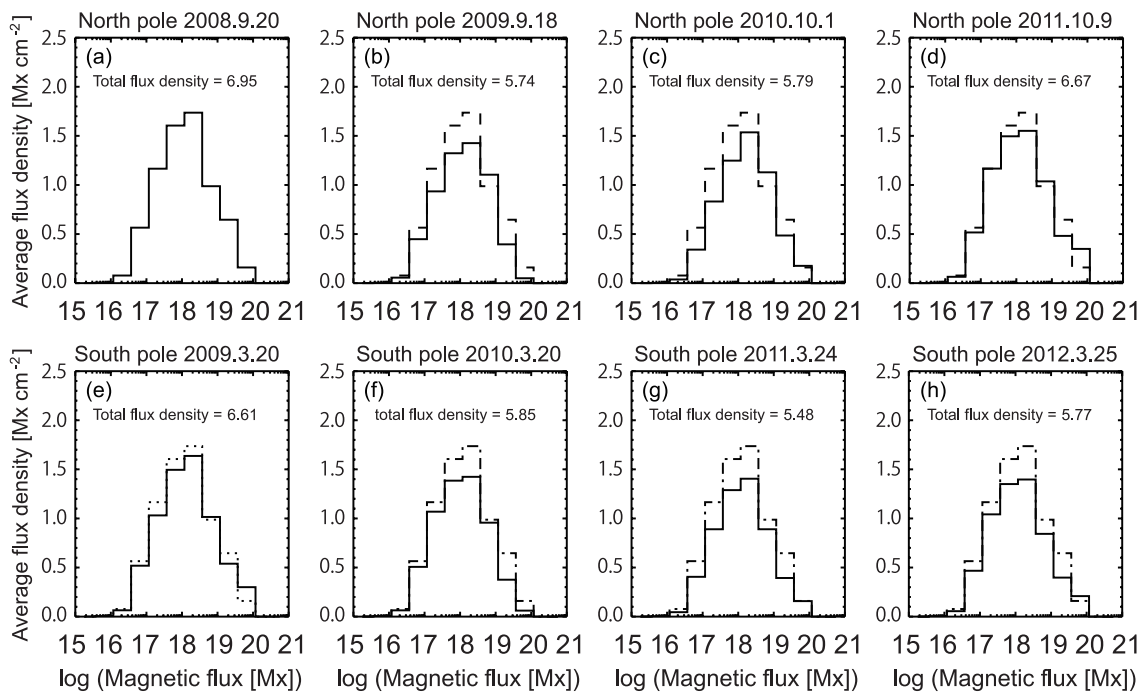}
\end{center}
\caption{
Histograms of magnetic flux density per patch in the north and south polar regions from 2008 to 2012. The magnetic flux concentrations here refer to the concentrations of the horizontal magnetic vectors. The average flux density is defined to be the total flux  (contributed by the magnetic concentrations in each bin of the horizontal axis) divided by the SOT observing area. Total average flux density is tabulated in each panel.  The observation dates are the same as those in Figure 3. The exposure times for these data are the same (12.8~s). The dashed lines in panels (b), (c), and (d) and the dotted line in panel (e) are the same as the histogram in panel (a),  and the dot-dashed lines in panels (f), (g), and (h) are the same as the histogram in panel (e) for comparison. 
}
\label{hf}
\end{figure}

\begin{figure}
\begin{center}
\includegraphics[height=12cm]{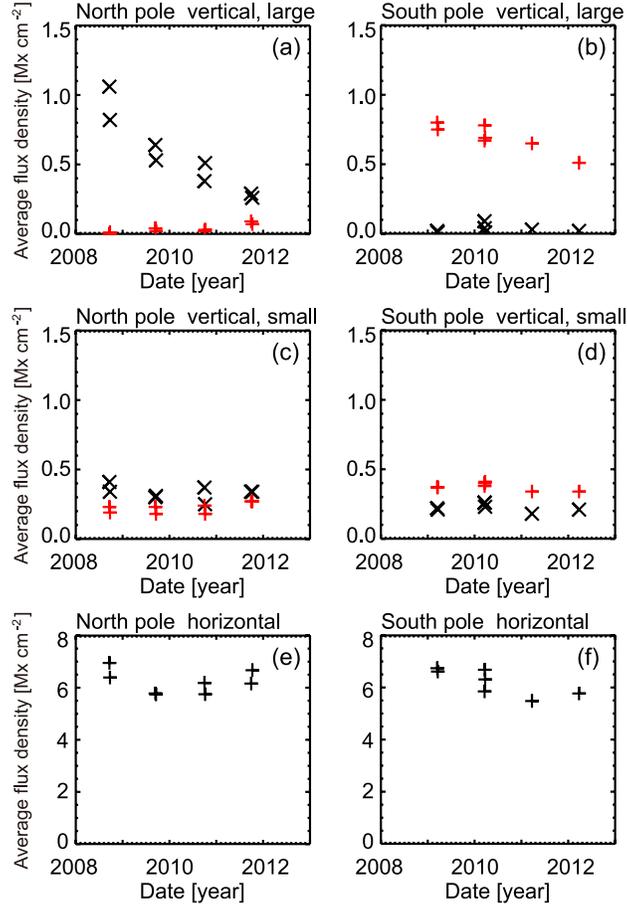}
\end{center}
\caption{
Yearly variation in the average flux density of the vertical and horizontal magnetic vectors in the north and south polar regions. Upper panels (a) and (b) show the average flux density of the vertical magnetic concentrations with total magnetic flux  (per patch) larger than  $ 10^{18}$~Mx . Red (cross) and black (oblique cross) symbols represent negative and positive polarities, respectively. Absolute values are shown in the panels. Middle panels (c) and (d) show the average flux density of the vertical magnetic concentrations with total magnetic flux (per patch) smaller than $ 10^{18}$~Mx. Bottom panels (e) and (f) show the average flux density of the horizontal magnetic vectors. All the average flux densities are numerically tabulated in Table 1.
}
\label{yv}
\end{figure}

\begin{figure}
\begin{center}
\includegraphics[width=12cm]{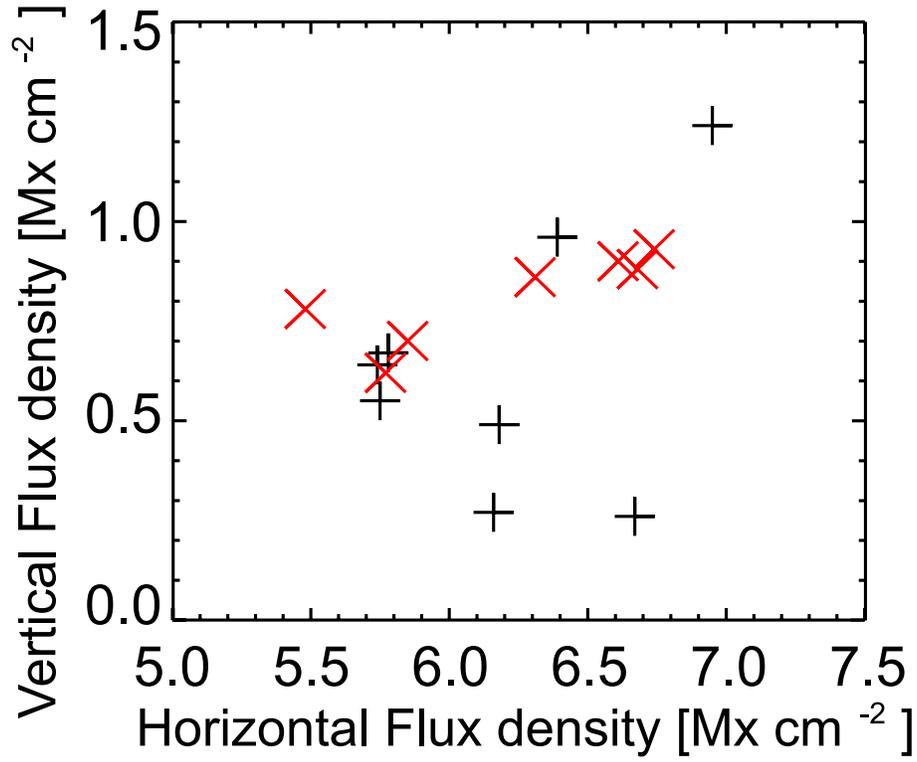}
\end{center}
\caption{
Scatter plot of the average flux density of the horizontal magnetic vectors  vs. the average flux density of the vertical magnetic vectors. Red (oblique cross) and black (cross) symbols represent south and north polar regions, respectively. 
}

\label{scatterplot}
\end{figure}

\begin{landscape}
\begin{table}
\scriptsize
\begin{center}
\caption{SOT/SP Polar Observations}
\label{datatable}
\begin{tabular}{llrrrrrrrrrrrrrr}

\hline 
& & \multicolumn{2}{c}{Time } & \multicolumn{1}{l}{ B0} & 
\multicolumn{4}{c}{Field of view \tablenotemark{a)}}  & Exp. & Polarization\tablenotemark{b)} & 
\multicolumn{5}{c}{Average flux density } 
 \\
  ID & \multicolumn{1}{c}{Date}  & Start & End &  \multicolumn{1}{l}{angle} &
$x_{\rm min}$\tablenotemark{ c)} & $x_{\rm max}$\tablenotemark{ c)} &
$y_{\rm min}$\tablenotemark{ c)} & $y_{\rm max}$\tablenotemark{ c)} &
time & mode & 
\multicolumn{1}{c}{$B_{\rm S,+}$ \tablenotemark{ d)} } & 
\multicolumn{1}{c}{$B_{\rm S,-}$ \tablenotemark{ d)} } & 
\multicolumn{1}{c}{$B_{\rm L,+}$ \tablenotemark{ e)} }&
\multicolumn{1}{c}{$B_{\rm L,-}$ \tablenotemark{ e)} }& 
\multicolumn{1}{c}{$B_{\rm h}$  \tablenotemark{ f)} } \\
 & & {\scriptsize(UT)} & {\scriptsize(UT)} & {\tiny(degree)} & 
{\tiny(arcsec)} & {\tiny(arcsec)} & {\tiny(arcsec)} & {\tiny(arcsec)} & {\scriptsize (s)} & &
{\tiny(Mx cm$^{\tiny-2}$)} & {\tiny(Mx cm$^{\tiny-2}$)} & {\tiny(Mx cm$^{\tiny-2}$)} &
{\tiny(Mx cm$^{\tiny-2}$)} & {\tiny(Mx cm$^{\tiny-2}$)}  \\

\hline 
N1 & 2008 Sep 20 & 10:12 & 17:38 &   7.09 &  
 -180 &   122 &   826 &   990 & 12.8 & 1 & 
   0.23 &   -0.41 &    0.00 &   -1.06 &    6.95 \\ 
 
N2 & 2008 Sep 24 & 10:37 & 17:43 &   6.97 &  
 -193 &   103 &   825 &   989 & 12.8 & 1 & 
   0.19 &   -0.34 &    0.01 &   -0.82 &    6.39 \\ 
 
N3 & 2009 Sep 13 & 10:41 & 16:45 &   7.22 & 
 -175 &    64 &   840 &  1004 & 12.8 & 1 & 
   0.23 &   -0.30 &    0.04 &   -0.64 &    5.78 \\ 
 
N4 & 2009 Sep 18 & 10:35 & 17:40 &   7.14 &  
 -166 &   129 &   843 &  1007 & 12.8 & 1 & 
   0.18 &   -0.31 &    0.02 &   -0.53 &    5.74 \\ 
 
N5 & 2010 Oct 01 & 10:05 & 17:30 &   6.71 & 
 -220 &    94 &   851 &  1015 & 12.8 & 1 & 
   0.24 &   -0.37 &    0.02 &   -0.38 &    6.18 \\ 
 
N6 & 2010 Oct 06 & 10:06 & 15:59 &   6.45 &  
 -204 &    39 &   853 &  1017 & 12.8 & 1 & 
   0.18 &   -0.25 &    0.03 &   -0.51 &    5.75 \\ 
 
N7 & 2011 Sep 30 & 10:04 & 15:30 &   6.77 &  
 -195 &    27 &   814 &   978 & 12.8 & 1 & 
   0.27 &   -0.34 &    0.09 &   -0.29 &    6.16 \\ 
 
N8 & 2011 Oct 09 & 10:04 & 17:30 &   6.28 &  
 -191 &   120 &   819 &   983 & 12.8 & 1 & 
   0.27 &   -0.34 &    0.07 &   -0.26 &    6.67 \\ 
\hline 
 
S1 & 2009 Mar 16 & 10:17 & 17:04 &  -7.14 &  
 -190 &    96 &  -999 &  -835 & 12.8 & 1 & 
   0.37 &   -0.22 &    0.80 &   -0.02 &    6.74 \\ 

S2 & 2009 Mar 20 & 08:01 & 13:51 &  -7.03 & 
 -159 &    85 &  -999 &  -836 & 12.8 & 1 & 
   0.37 &   -0.21 &    0.75 &   -0.01 &    6.61 \\ 

S3 & 2010 Mar 20 & 11:13 & 17:29 &  -7.04 &  
 -178 &    81 & -1002 &  -838 & 12.8 & 1 & 
   0.38 &   -0.26 &    0.67 &   -0.09 &    5.85 \\ 

S4 &  2010 Mar 21 & 10:07 & 17:44 &  -7.01 &  
 -193 &   100 & -1007 &  -843 & 12.8 & 1 & 
   0.40 &   -0.26 &    0.78 &   -0.04 &    6.68 \\ 

S5 &  2010 Mar 24 & 10:50 & 16:49 &  -6.90 &  
 -185 &    63 & -1000 &  -836 & 12.8 & 1 & 
   0.41 &   -0.23 &    0.69 &   -0.01 &    6.31 \\ 

S6 & 2011 Mar 24 & 11:04 & 17:09 &  -6.91 &  
 -173 &    79 &  -993 &  -829 & 12.8 & 1 & 
   0.34 &   -0.18 &    0.65 &   -0.03 &    5.48 \\ 

S7 & 2012 Mar 25 & 08:36 & 15:59 &  -6.85 &  
 -186 &   97 &  -999 &  -835 & 12.8 & 1 & 
   0.34 &   -0.21 &    0.51 &   -0.02 &    5.77 \\ 
\hline 

QS\tablenotemark{ g)} & 2007 Nov 28 & 18:36 & 20:52 &  1.22 &  
 -983 &  -883 &   -84 &    79 &  12.8 & 2& 
 -- & -- & -- & -- & -- \\

\hline 
\end{tabular}
\end{center}
\tablenotetext{a)}{\scriptsize  Field of view in heliographic coordinates}\tablenotetext{b)}{\scriptsize  one polarization channel (1) or two orthogonal polarization channels (2) are used.}
\tablenotetext{c)}{\scriptsize $x$-axis is directed toward solar west, and $y$-axis to solar north. The origin of the $x$--$y$ coordinate is the Sun center.}
\tablenotetext{d)}{\scriptsize  Average flux density of the vertical magnetic concentrations with total magnetic flux  smaller than  $ 10^{18}$~Mx.}
\tablenotetext{e)}{\scriptsize  Average flux density of the vertical magnetic concentrations with total magnetic flux   larger than  $ 10^{18}$~Mx.}
\tablenotetext{f)}{\scriptsize  Average flux density of the horizontal magnetic concentrations}
\tablenotetext{g)}{\scriptsize  Quiet Sun for reference: the same region observed 2.5 h after the resent data were acquired, which extensively analyzed by \cite{2010ApJ...719..131I}. The observation mode is different from the polar data, and so the magnetic flux density is not shown.}
\end{table}
\end{landscape}

\begin{figure}
\begin{center}
\includegraphics[width=12cm]{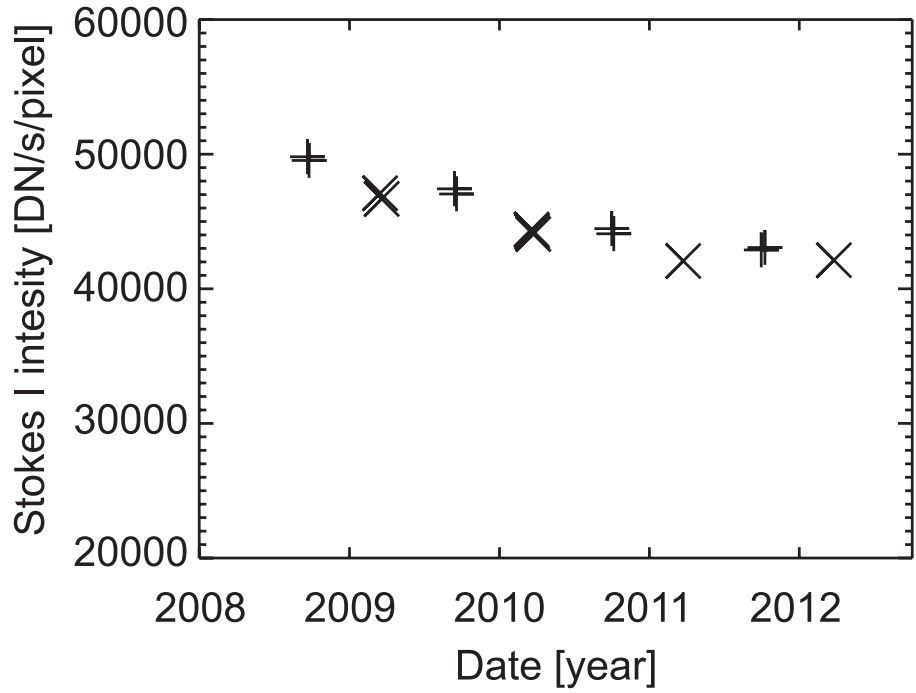}
\end{center}
\caption{
Long-term variation in the Stokes I continuum intensity as detected by SOT/SP. The data points correspond to those listed in Table 1. The average count rate for a small region fixed relative to the solar limb is calculated. Oblique cross and cross symbols represent the south and north polar regions, respectively. 
}
\label{Icont}
\end{figure}

\end{document}